\documentclass[]{pasj01}

\draft

\Received{$\langle$reception date$\rangle$}
\Accepted{$\langle$acception date$\rangle$}
\Published{$\langle$publication date$\rangle$}

\begin{document}

\title{The Contribution of Chemical Abundances in Nova Ejecta to the Interstellar Medium}
\author{Fanger Li\altaffilmark{1}, Chunhua Zhu\altaffilmark{1},  Guoliang L\"{u}\altaffilmark{1}, Zhaojun Wang\altaffilmark{1}}
\altaffiltext{1}{School of Physical Science and Technology, Xinjiang
University, Urumqi, 830046, China}
\email{1442796381@qq.com, xjdxwzj@sohu.com}


\maketitle


\begin{abstract}
According to the nova model from \citet{Yaron2005} and \citet{Jose1998} and using
Monte Carlo simulation method,
we investigate the contribution of chemical abundances in nova ejecta
to the interstellar medium (ISM) of the Galaxy. We find that the ejected mass by classical novae (CNe) is about
$2.7\times10^{-3}$ $ \rm M_\odot\ {\rm yr^{-1}}$. In the nova ejecta, the isotopic ratios of C, N and O,
that is, $^{13}$C/$^{12}$C, $^{15}$N/$^{14}$N and $^{17}$O/$^{16}$O,
are higher about one order of magnitude than those in red giants.
We estimate that about 10$\%$, 5$\%$ and 20$\%$ of $^{13}$C, $^{15}$N and $^{17}$O
in the ISM of the Galaxy come from nova ejecta, respectively.
However, the chemical abundances of
C, N and O calculated by our model can not cover all of observational values.
This means that there is still a long way to go for understanding novae.
\end{abstract}

\section{Introduction}
Chemical elements are the significant basic components of the interstellar medium (ISM),  which
includes gas in ionic, atomic, and molecular, as well as dust and cosmic rays.
They play an active role
in the astrophysics and chemical evolution of the ISM
and participate in a cycle of the Galactic chemical evolution
in which synthesis of heavy elements, gas and dust grains in the ejecta of evolved stars are a crucial reservoir
for the transportation of metals in the Galactic eco-system.
In the diffuse gas, the materials are eventually incorporated into
young stars and planetary systems during star formation.
According to popular point of view, low and intermediate mass stars during the Asymptotic Giant Branch (AGB) stage of their evolution,
supernovae and Wolf-Rayet stars, have long been considered
the main sites of contributors to the ISM (at least on local scales)
of highly processed materials \citep{Gail2009,Dunne2003}.

The observations of elemental enrichments of the Galactic classical novae (CNe) indicate that
the role of novae has been received considerable attention
and novae are
the major producers of the freshly produced elements \citep{Truran1985}.
The overall mass contribution from novae to the ISM is quite small
that novae probably have processed less than $\sim 0.3\%$ of the ISM in the Galaxy \citep{Gehrz1998},
but both observational and theoretical evidences imply that novae may be the major sources of the odd-numbered nuclei $^{13}$C, $^{15}$N and $^{17}$O,
which differ markedly from solar,
and novae clearly contribute to the chemical element evolution of the Galaxy \citep{Starrfield1997,Timmes1997,Jose1998,Gehrz1998,Hix2001,Jose2006}.
Such nuclei are difficult to form in other astrophysical hosts.

Nova outbursts are the results of thermonuclear runaways (TNRs) of hydrogen on the surfaces of white dwarfs (WDs) components of
Cataclysmic Variable (CV) systems.
The companions in close binary systems expand beyond their Roche lobes and
pass matter via the inner Lagrangian point, leading to the formation of a accretion disk around the WD \citep{Starrfield1972,Gehrz1998,Bode2008}.
Subsequently,
accumulation of the hydrogen-rich envelope on the WD continues until the critical explosive conditions are achieved
because of compressional heating and energy release from nuclear reaction at the base of the accreted envelope and a TNR ensues,
leading to a violent explosion of materials \citep{Jose2006,Starrfield2008}.
Both observational and theoretical studies of the consequences of the TNR strongly suggest that, at some point,
the matter from the underlying WD is mixed with the accreted matter,
and this mixing is heated and compressed, leading eventually to inject plentiful materials nuclear-processed
by explosive hydrogen-rich burning nucleosynthesis into the ISM,
but when and how the mixture occurs is still debated.
Neither the mechanism responsible for the mixing nor the phase of the outburst
in which the mixing occurs is universally accepted \citep{Gehrz1998,Yaron2005,Starrfield2008,Shara2010}.
The influence of the chemical abundances in nova ejecta on the ISM is generally global.
Therefore, we concentrate specifically on the gas-phase chemical anomalies by novae that may have been contributed to the ISM of the Galactic.

In short, analyzing the enrichments of chemical elements in the ejected materials
are the key factors for understanding the chemical evolution of galaxy and are
very interested and excited.
A series of observational data and theoretical works of novae have been published, and they show important overproduction of
several nuclides, such as
$^{13}$C, $^{15}$N and $^{17}$O
\citep{Starrfield1972,Sparks1976,Prialnik1986,Politano1995,Kovetz1997,Starrfield1997,Jose1997,Timmes1997}.
Particularly, Galactic $^{15}$N and $^{17}$O have been highly supported as being produced during nova eruptions \citep {Woosley1997}, and for the production of $^{13}$C isotopes,
a low value for the ratio $^{12}$C/$^{13}$C = 5 was reported by \citet{Pavlenko2004} in V4334 Sgr.

\citet{Schild1992} reported the chemical abundances of the ejected materials in several novae.
\citet{Downen2012} and \citet{Downen2013} showed how useful the elemental abundances are for constraining the peak temperature achieved during the CN outbursts.
In particular, \citet{Kovetz1997}, \citet{Jose1998} and \citet{Yaron2005} carried out detailed numerical simulations
for the chemical abundances of the novae with carbon-oxygen (CO) WD or oxygen-neon-magnesium (ONeMg) WD accretors.
This makes it possible to simulate the elemental abundances of the ejecta by Mente Carlo technology.

Of particular important to the study, in this paper, is that the abundances of the chemical elements in nova ejecta
and their contributions to the ISM
are carried out via a population synthesis method.
In the next section, we present the model that are used for the reported calculations in this paper.
In the following section, we describe the Monte Carlo simulation technique.
In Section 4, we show the results of our new calculations and discussion, and
in Section 5, we continue with a summary of the results.

\section{Nova model and input physics}
The rapid binary star evolution (BSE) code \citep{Hurley2002}
is employed here for the simulation of binary evolutions.
Below, we mention the updated model used in our code.

\subsection{Nova model}
Following the work described in \citet{Yaron2005} and \citet {Jose1998}, a updated nova evolution code is developed
in some details
to analyze the behaviors of nova explosions
from the onset of accretion up to ejection stages.
It is multiple nova evolution model that
the different characteristics of nova eruptions can be produced by controlling three basic independent parameters
: the WD mass, the temperature of its isothermal core
and the mass accretion rate from the companion.
In addition,
we assume that the material being accreted from the donor star is of Solar composition \citep{Anders1989},
and that it has already mixed with the core materials, such that
the actual accreting ingredients is approximately $50\%$ solar and $50\%$ WD material for all the models in this study \citep{Politano1995}.

Novae phenomena are produced by TNRs on the surfaces of WDs accreting hydrogen matter in the close binary systems.
The range of possible mass accretion rates is very wide.
Different extents of nova outburst is expected to occur for different mass accretion rates.
It should stays in one of the following three phases:
(i) The stable hydrogen burning phase at the surface for accretion rates between $1.03\times 10^{-7}$ $ \rm M_\odot$ $ \rm yr^{-1}$ and $2.71\times 10^{-7} $ $\rm M_\odot$ $ \rm yr^{-1}$ \citep{Hurley2002};
(ii) The TNR phase for mass accretion rates between $5.0\times 10^{-13}$ $ \rm M_\odot$ $ \rm yr^{-1}$ and $1.03\times 10^{-7}$ $ \rm M_\odot$ $ \rm yr^{-1}$ \citep{Yaron2005};
(iii) The declining phase after a thermonuclear eruption for rates lower than $5.0\times 10^{-13}$ $ \rm M_\odot$ $ \rm yr^{-1}$.

The efficiency of mass accumulation by a WD can never be 100$\%$.
Firstly,
it strongly relys on the strength of novae;
secondly, even steady burning WDs blow stellar wind.
In present paper, we accept the prescription of \citet {lv2006}.
The strength of a nova outburst depends on the mass and mass accretion rate of the WD.
According to \citet{Yaron2005}, novae are roughly separated into strong novae and weak novae
via the mass accretion rate $\dot{ M }_{\rm ws}$ (in $ \rm M_\odot$ $ \rm yr^{-1}$)
\begin{eqnarray}&&
\log_{10} \dot{M}_{\rm ws}=\left\{
\begin{array}{cc}
-11.01+6M_{\rm WD}-1.90{M}^2_{\rm WD},\ \ & M_{\rm WD} \leq 1\rm M_\odot;\\
-7.0 ,\ \ & M_{\rm WD} > 1\rm M_\odot.
\end{array}\right.
\end {eqnarray}
which is given by \citet{lv2006}.
For weak novae,
mass-accretion rates are between a critical rate $\dot{ M }_{\rm ws}$ and
$1.03\times 10^{-7}$ $ \rm M_\odot$ $ \rm yr^{-1}$;
and for strong novae,
mass-accretion rates are lower than  $\dot{ M }_{\rm ws}$, a large proportion of the accreted materials are blown away
and even in some case an erosion of the WD occurs for strong novae \citep{Prialnik1995,Yaron2005}.

\subsection{Chemical abundances}
Before describing our simulations, we summarize all the investigated abundances which have observed in nova shells.
Compilations of CN abundances (expressed as the mass fractions) can be found, for example, in
\citet {Livio1994}, \citet {Starrfield1997}, \citet {Gehrz1998}, \citet {Wanajo1999} and \citet {Downen2013}.
Novae with WD accretors eject materials into the ISM.
A detailed study on the chemical abundances of novae had been showed by \citet {Kovetz1997} and \citet {Jose1998} during the thermonuclear outbursts.
In this paper, we
present the elemental abundances of the ejected materials during the thermonuclear outbursts.

CNe may be the only objects to observe directly all aspects in ejecta on a regular basis,
which can provide
significant information of the evolution and characteristics of the observed systems.
The CNe last several days or weeks and the variation of their visual magnitudes range from -5 to -10.
Based on a number of typical observations,
nova events are usually divided into two fundamentally different types in the light of the composition of progenitor stars: CO and ONeMg novae
\citep {Bode1989,Gehrz1995,Jose1998}.

Numerical calculations of the nova explosions
have been performed by \citet {Shara1993}, \citet {Kovetz1994}, \citet {Prialnik1995}
and \citet {Yaron2005}.
\citet {Kovetz1997} carried out detailed multicycle calculations of the nova eruptions for CO WDs with masses ranging from 0.65 to 1.4 $\rm M_\odot$.
In their work,
the element abundances of nova ejecta rely on four independent and basic parameters:
the mass accretion rate $\dot{M}_{\rm WD}$, the WD mass $M_{\rm WD}$, its core temperature $ T_{\rm WD} $ and the C/O ratio of the accreting WD.
In order to confirm the effect of the WD composition on the abundances of nova ejecta,
\citet{Kovetz1997} calculated the nova models with WD accretors consisted of pure C, pure O and C/O=1,
and they found the WD composition is not reflected in the abundances of the ejecta.
Furthermore,
according to the work of \citet {Iben1985}, the AGB phase may be suppressed in a close binary system
and a formed WD should have a ratio very close to unity in an extensive study of close binary evolution.
Hence, in this work, we assume that the model have an identical homogeneous composition of C and O in equal mass fraction for CO WDs.
Neglecting the effect of the nova explosions on the temperature of the accreting WD.
For the abundances of $^{1}$H and $^{4}$He, and the abundances of $^{12}$C, $^{13}$C, $^{14}$N, $^{15}$N, $^{16}$O and $^{17}$O, we select 26 models in \citet{Yaron2005} and 10 models in \citet {Kovetz1997} ( see Table \ref{tab:hhe} and Table \ref{tab:cno}, where $Z_{\rm ej} $ is the total heavy element of ejecta.), respectively, in which $ T_{\rm WD} $= $3\times10^{7}$ K  and $\rm C/O$ ratio of the accreting WD is 1.
Their abundances are determined by the mass-accretion rate  $\dot{M}_{\rm WD}$ and the mass $M_{\rm WD}$.
By a bilinear interpolation \citep{Press1992} of 10 models in \citet {Kovetz1997} and 26 models in \citet{Yaron2005},
the abundances of $^{1}$H, $^{4}$He, $^{12}$C, $^{13}$C, $^{14}$N, $^{15}$N, $^{16}$O  and $^{17}$O
are calculated in the nova ejecta.
If $\dot{M}_{\rm WD}$ or $M_{\rm WD}$ values in systems are not in the range of the bilinear interpolation, they are taken as the most vicinal in models.

The abundances of $^{20}$Ne and $^{22}$Ne can not be found in \citet{Kovetz1997}.
\citet{Jose1998} gave the nucleosynthesis in the nova eruptions with CO and ONeMg WD accretors.
In their work, the nova outbursts are affected by $M_{\rm WD}$, $\dot{M}_{\rm WD}$, the initial luminosity (or $T_{\rm WD}$),
and the mixing level between core and envelope.
In order to verify the influence of the WD mass,
they followed a series of simulations involving CO WDs ($M_{\rm WD}$=0.8, 1.0, and 1.15 $\rm M_\odot$ )
and ONeMg WDs ($M_{\rm WD}$=1.0, 1.15, 1.25, and 1.35 $\rm M_\odot$ ). The mass accreting rate is $2\times10^{10}$ $\rm M_\odot$ $ \rm yr^{-1}$,
and the initial luminosity is 10$^{-2} $
$\rm L_\odot$.
The mixability between the core and the envelope is a special uncertain parameter.
\citet{Jose1998} modeled three different mixing levels: 25$\%$, 50$\%$ and 75$\%$.
Following \citet{Starrfield1998}, a 50$\%$ degree of mixing is taken here.
Furthermore, we adopt the fitting formulae of $^{20}$Ne and  $^{22}$Ne in \citet{Lv2008} to calculate their abundances for CO novae,
which selected
the three nova models (the mixing level is 50$\%$) in \citet{Jose1998} ( see Table \ref{tab:nne} ).
In the above fitting formulaes, the fractions of $^{20}$Ne and  $^{22}$Ne depend weakly on $M_{\rm WD}$.

\begin{table}
\footnotesize
\caption{
The mass fractions of $^{1}$H $(X_{\rm ej})$, $^{4}$He $(Y_{\rm ej})$ and the heavy element $(Z_{\rm ej} )$ of ejecta in CO nova model \citep {Yaron2005}, in which $ T_{\rm WD} $= $3\times10^{7}$ K and $\rm C/O$ ratio of the accreting WD is 1.
}
\begin{center}
\begin{tabular}{lllllllllll} 
\hline\hline
$M_{\rm WD}$ ($ \rm M_\odot$) &  $\rm log \dot{ M }_{\rm WD}$ ($ \rm M_\odot$ $ \rm yr^{-1}$) &
$X_{\rm ej}$ & $Y_{\rm ej}$ & $Z_{\rm ej} $\\
\hline
       0.65 \ \ &   -8    \ \ &   0.5985  \ \ &    0.3801  \ \ &   0.0214  \ \ &\\
            \ \ &   -9    \ \ &   0.5942  \ \ &    0.2956  \ \ &   0.1102  \ \ &\\
            \ \ &   -10   \ \ &   0.5141  \ \ &    0.2489  \ \ &   0.2370  \ \ &\\
            \ \ &   -11   \ \ &   0.4273  \ \ &    0.1817  \ \ &   0.3910  \ \ &\\
            \ \ &   -12   \ \ &   0.2699  \ \ &    0.1216  \ \ &   0.6085  \ \ &\\
            \ \ &   -12.3 \ \ &   0.2539  \ \ &    0.1161  \ \ &   0.6300  \ \ &\\
       1.00 \ \ &   -7    \ \ &   0.6250  \ \ &    0.3535  \ \ &   0.0215  \ \ &\\
            \ \ &   -8    \ \ &   0.5690  \ \ &    0.3300  \ \ &   0.1010  \ \ &\\
            \ \ &   -9    \ \ &   0.5470  \ \ &    0.2940  \ \ &   0.1590  \ \ &\\
            \ \ &   -10   \ \ &   0.4855  \ \ &    0.2587  \ \ &   0.2558  \ \ &\\
            \ \ &   -11   \ \ &   0.3460  \ \ &    0.1884  \ \ &   0.4656  \ \ &\\
            \ \ &   -12   \ \ &   0.2364  \ \ &    0.1665  \ \ &   0.5971  \ \ &\\
            \ \ &   -12.3 \ \ &   0.2505  \ \ &    0.2064  \ \ &   0.5431  \ \ &\\
       1.25 \ \ &   -7    \ \ &   0.5347  \ \ &    0.4380  \ \ &   0.0273  \ \ &\\
            \ \ &   -8    \ \ &   0.5292  \ \ &    0.3671  \ \ &   0.1037  \ \ &\\
            \ \ &   -9    \ \ &   0.5241  \ \ &    0.3400  \ \ &   0.1359  \ \ &\\
            \ \ &   -10   \ \ &   0.4460  \ \ &    0.3060  \ \ &   0.2480  \ \ &\\
            \ \ &   -11   \ \ &   0.2867  \ \ &    0.2202  \ \ &   0.4931  \ \ &\\
            \ \ &   -12   \ \ &   0.2151  \ \ &    0.2124  \ \ &   0.5725  \ \ &\\
            \ \ &   -12.3 \ \ &   0.1891  \ \ &    0.2708  \ \ &   0.5401  \ \ &\\
       1.40 \ \ &   -7    \ \ &   0.5155  \ \ &    0.4574  \ \ &   0.0271  \ \ &\\
            \ \ &   -8    \ \ &   0.4251  \ \ &    0.5410  \ \ &   0.0339  \ \ &\\
            \ \ &   -9    \ \ &   0.3612  \ \ &    0.4891  \ \ &   0.1497  \ \ &\\
            \ \ &   -10   \ \ &   0.3032  \ \ &    0.4643  \ \ &   0.2325  \ \ &\\
            \ \ &   -11   \ \ &   0.1689  \ \ &    0.3986  \ \ &   0.4325  \ \ &\\
            \ \ &   -12   \ \ &   0.0608  \ \ &    0.3213  \ \ &   0.6179  \ \ &\\
\hline \hline
\end{tabular}
\end{center} \label{tab:hhe}
\end{table}

\begin{table}
\footnotesize
\caption{
The mass fractions of CNO isotope in CO nova ejecta \citep{Kovetz1997}, in which $ T_{\rm WD} $= $3\times10^{7}$ K  and $\rm C/O$ ratio of the accreting WD is 1.
}
\begin{center}
\begin{tabular}{lllllllllll} 
\hline\hline
$M_{\rm WD}$ &  $\rm log \dot{ M }_{\rm WD}$ \\
($ \rm M_\odot$) & ($ \rm M_\odot$ $ \rm yr^{-1}$) &
$^{12}$C &  $^{13}$C &
$^{14}$N &  $^{15}$N &  $^{16}$O  & $^{17}$O \\
\hline
       0.65 \ \ &   -9  \ \ &   2.24E-03  \ \ &   7.41E-04  \ \ &   5.63E-02  \ \ &   2.38E-06  \ \ &   4.81E-02  \ \ &   2.86E-03  \ \ &\\
            \ \ &   -10 \ \ &   9.28E-03  \ \ &   5.09E-03  \ \ &   1.11E-01  \ \ &   1.12E-05  \ \ &   1.05E-01  \ \ &   3.31E-03  \ \ &\\
       1.00 \ \ &   -8  \ \ &   2.69E-03  \ \ &   9.14E-04  \ \ &   5.79E-02  \ \ &   4.26E-06  \ \ &   3.57E-02  \ \ &   4.21E-03  \ \ &\\
            \ \ &   -9  \ \ &   4.39E-03  \ \ &   1.68E-03  \ \ &   7.38E-02  \ \ &   7.96E-06  \ \ &   5.19E-02  \ \ &   5.47E-03  \ \ &\\
            \ \ &   -10 \ \ &   1.15E-02  \ \ &   7.82E-03  \ \ &   1.21E-01  \ \ &   1.89E-04  \ \ &   1.07E-01  \ \ &   7.97E-03  \ \ &\\
       1.25 \ \ &   -8  \ \ &   4.38E-03  \ \ &   1.59E-03  \ \ &   7.35E-02  \ \ &   1.61E-05  \ \ &   2.18E-02  \ \ &   2.62E-03  \ \ &\\
            \ \ &   -9  \ \ &   8.17E-03  \ \ &   4.04E-03  \ \ &   9.37E-02  \ \ &   1.88E-04  \ \ &   2.45E-02  \ \ &   3.92E-03  \ \ &\\
            \ \ &   -10 \ \ &   2.20E-02  \ \ &   1.66E-02  \ \ &   1.27E-01  \ \ &   1.40E-03  \ \ &   7.84E-02  \ \ &   1.35E-02  \ \ &\\
       1.40 \ \ &   -9  \ \ &   2.10E-02  \ \ &   1.34E-02  \ \ &   1.17E-01  \ \ &   1.27E-03  \ \ &   2.35E-04  \ \ &   1.51E-05  \ \ &\\
            \ \ &   -10 \ \ &   3.74E-02  \ \ &   3.15E-02  \ \ &   1.58E-01  \ \ &   4.92E-03  \ \ &   3.12E-04  \ \ &   4.87E-05  \ \ &\\
\hline \hline
\end{tabular}
\end{center} \label{tab:cno}
\end{table}

\begin{table}
\footnotesize
\caption{
The mass fractions of $^{20}$Ne and $^{22}$Ne in CO nova ejecta \citep{Jose1998}, in which the mass accreting rate is $2\times10^{10}$ $\rm M_\odot$ $ \rm yr^{-1}$, the initial luminosity is 10$^{-2} $
$\rm L_\odot$ and the mixing level is 50$\%$.
}
\begin{center}
\begin{tabular}{lllllllllll} 
\hline\hline
$M_{\rm WD}$ &  $^{20}$Ne  & $^{22}$Ne \\
\hline
       0.80 \ \ &   8.2E-04  \ \ &   5.0E-03  \ \ &\\

       1.00 \ \ &   8.5E-04  \ \ &   5.0E-03  \ \ &\\

       1.15 \ \ &   9.7E-04  \ \ &   4.8E-03  \ \ &\\
\hline \hline
\end{tabular}
\end{center} \label{tab:nne}
\end{table}

For the ONeMg novae, \citet {Lv2008} gave the fitting formulae of the element abundances of $^{1}$H, $^{4}$He, $^{12}$C, $^{13}$C, $^{14}$N, $^{15}$N, $^{16}$O, $^{17}$O $^{20}$Ne and $^{22}$Ne (see formulae (8) in \citet {Lv2008}) by fitting data in \citet{Jose1998}
in which the mixing level of the selected models is 50$\%$,
and the formulaes agree with the simulated results to within a factor of 1.3.
We ignore other chemical elements on account of their abundances smaller than the above element abundances or their isotopes.
All elemental abundances in the ejecta are renormalized and their sum is 1.0.

\section{Basic parameters of the Monte Carlo simulation}
Population synthesis is to evolve large numbers of stars in order to
investigate and understand statistical properties of stars.
Here,
for purpose of investigating the Galactic occurrence rate of novae and the contribution of chemical abundances in nova ejecta
to the ISM of the Galaxy, we perform a Monte Carlo simulation for a sample of $10^{6}$ binary systems.

For the population synthesis of binary systems, the main input model parameters
require the star formation rate (SFR), the initial mass function (IMF) of the primaries,
the initial mass ratio distribution of binaries, the distribution of initial orbital separations,
the eccentricity distribution and the metallicity $Z$ of the binary systems.

(1) The SFR is taken to be constant over the last 13 \rm G$\rm yr$.

(2) A simple approximation to the IMF of \citet{Miller1979} is used;
the primary mass is generated with the formula of \citet{Eggleton1989}:
\begin{equation}
M_1=\frac{0.19X}{(1-X)^{0.75}+0.032(1-X)^{0.25}}
\end{equation}
where $X$ is a random number uniformly distributed between 0 and 1. The
adopted masses of primaries are more than 0.8 $\rm M_\odot$.

(3) The mass ratio distribution is quite controversial. We mainly take a constant mass ratio distribution \citep{Mazeh1992,Goldberg1994},
\begin{equation}
n(q)=1,  0<q\leq 1,
\end{equation}
where  $ q=M_2/M_1$.

(4) The distribution of orbital separations is given by
 $\rm log$ $ a = 5X + 1$
, where $X$ is a random variable uniformly distributed in the range
[0,1] and $a$ is in $ \rm R_\odot$ \citep{Yungelson1993,lv2006}.

(5) The metallicity $Z$ is 0.02 of Population I \citep{Iben1966,Kippenhahn1990}, and all binary systems have initially circular orbits.

We follow the evolutions of both components with the rapid BSE code
including the effect of tides on binary evolution \citep {Hurley2002}.
This code contains many characteristics of binaries, that is, mass transfer, mass accretion, common-envelope evolution, collisions, supernova kicks and angular momentum loss mechanism, etc.
Another model parameters in the population synthesis are common envelope (CE) ejection efficiency $\alpha _{\rm CE}$
and the binding energy factor  $\lambda $  for CE evolution.
The CE ejection efficiency $\alpha _{\rm CE}$ is the fraction of the released orbital energy used to
overcome the binding energy of the envelope during the spiral-in process of a CE.
It is probably not a constant \citep {Regos1995}, but generally a typical value for the parameter $\alpha _{\rm CE}$ is 1.0, that is, all the released orbital energy are transferred to
the envelope to overcome its binding energy.
The parameter $\lambda $, which is used in calculations of the envelope binding energy for giants in CE, is taken as 0.5 \citep {Hurley2002}.
In this work, if we do not particularly mention for input parameter, it is taken as the default value in \citet {Hurley2002}.

\section{Results and discussions}
We construct model to analyze the final abundances of the chemical elements in nova ejecta.
We
also present the final isotopic abundances ratios.

\subsection{Galactic occurrence rate of novae}

In order to investigate the significant occurrence rate and properties of novae, we
follow the evolutions of
$10^{6}$ initial binary systems for simulation via a Monte Carlo simulation method.
Meanwhile,
we support that a binary with its primary more massive than 0.8 $\rm M_\odot$ is formed annually in the
Galaxy \citep{Phillips1989,Yungelson1994,Han2004, Zhang2005}.
For simulations with $10^{6}$ binaries in our work, the statistical error of the numbers of nova outbursts is less than 1\%.
Thus, $10^{6}$ initial binaries seem to be an acceptable sample for our study.

Theoretical studies indicate that the average mass returned by a CN explosion
to the ISM is $\sim 2\times10^{-4}$ $\rm M_\odot\ {\rm yr^{-1}}$ \citep {Gehrz1998}.
Based on observational data,  \citet{Shafter1997} estimated the occurrence rate of CNe is (35$\pm$11) ${\rm yr^{-1}}$ in our Galaxy.
This means that CNe eject $\sim7\times10^{-3}$ $\rm M_\odot\ {\rm yr^{-1}}$ of the processed material into the ISM.
In our simulation,
the occurrence rate of CN explosions is about 54 $\rm yr^{-1}$.
It is larger than the observations because
not all novae can be observed.
And the average ejected mass per event is $\sim5.0\times10^{-5}$ $\rm M_\odot$,
the total amount of the ejecta is $\sim2.7\times10^{-3} $ $\rm M_\odot\ {\rm yr^{-1}}$. Our result is close to the observational value.

\subsection{ CNO abundances}
We calculate the chemical abundances of C, N and O elements in the nova ejecta.
However, it is very difficult to directly measure their abundances. Observationally,
the chemical abundance ratios can be observed.
\citet{Vogel1992} gave the C/N and O/N abundances ratios in the early nebular phase for
the eruption of PU Vul in 1977. Then, \citet{Livio1994}, \citet{Prialnik1997}, \citet{Prialnik1998},
\citet{Starrfield1997}, \citet{Gehrz1998} and \citet{Downen2013} measured the chemical abundance ratios
of many CNe.

\begin{figure}
 \begin{center}
\includegraphics[totalheight=3.0in,width=2.2in,angle=-90]{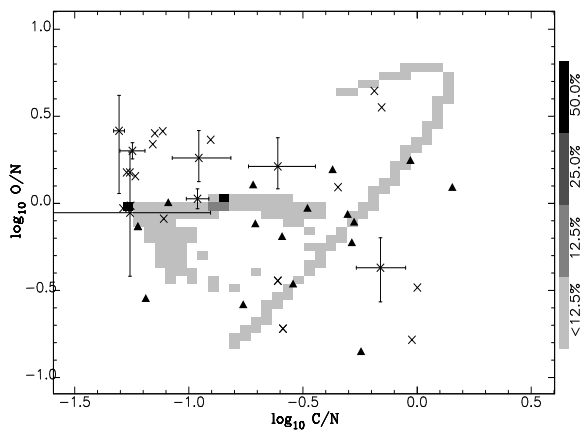}
 \end{center}
\caption{Gray-scale plot of the elemental abundance ratios of \rm log O/N vs. \rm log C/N for CO and ONeMg novae.
The left region represents novae with CO WD accretors
and the right linear region represents novae with ONeMg WD accretors.
The filled triangles and the crosses represent the observed values of CO and ONeMg novae, respectively.
In this paper, all of the observed results are
from \citet{Livio1994}, \citet{Prialnik1997}, \citet{Prialnik1998}, \citet{Starrfield1997}, \citet{Gehrz1998} and \citet{Downen2013}.
}
\label{fig:g-cno}
\end{figure}

Figure \ref{fig:g-cno} shows the distributions of C/N vs. O/N of CNe observed and our simulations.
The filled triangles and the crosses represent the observed values of CO novae and ONeMg novae, respectively.
It is very evident that the distributions of C/N vs. O/N in our results are divided into two regions.
The left region represents
novae with CO WD accretors, and the right linear region manifests novae with ONeMg WD accretors.
Unfortunately, our results do not cover most of CNe observed.
We consider that there are the following reasons:
First, the nova model used in this work is too rough, especially, the formula fitted by \citet{Lv2008} and the data
in \citet{Kovetz1997} are not suitable for all mass-accretion rates.
Second, there are very errors in measuring the chemical abundance ratios of CNe. As \citet{Downen2013} mentioned,
the values observed by different authors, for identical nova, are pretty different. Sometimes,
these different may come from different authors. Therefore, not only observationally but also theoretically, our understanding for
novae is very poor.

\begin{figure}
 \begin{center}
\includegraphics[totalheight=3.0in,width=2.2in,angle=-90]{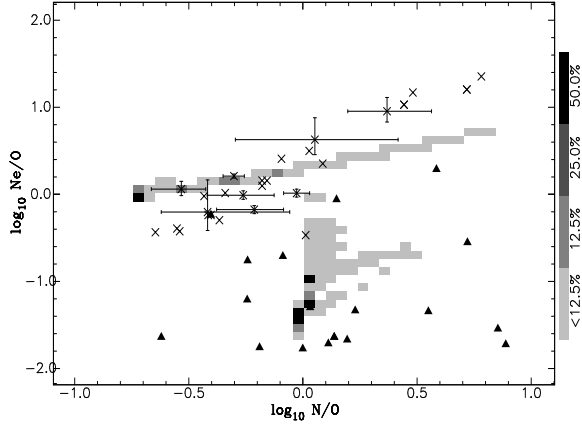}
 \end{center}
\caption{Gray-scale map of the elemental abundance ratios of \rm log Ne/O vs. \rm log N/O for CO and ONeMg novae.
The bottom region represents novae with CO WD accretors
and the top region represents novae with ONeMg WD accretors.
We use the same notations for the observed results as in Figure \ref{fig:g-cno}.
}
\label{fig:g-nneo}
\end{figure}

\begin{figure}
 \begin{center}
\includegraphics[totalheight=3.0in,width=2.2in,angle=-90]{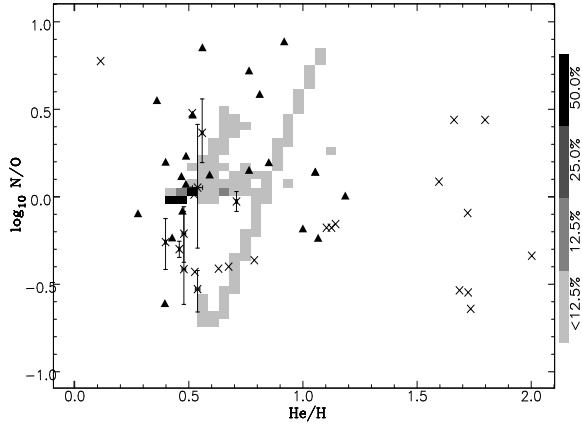}
 \end{center}
\caption{Gray-scale map of the elemental abundance ratios of \rm log N/O vs. He/H for CO and ONeMg novae.
The left region represents novae with CO WD accretors
and the right linear region manifests novae with ONeMg WD accretors.
We use the same notations for the observed results as in Figure \ref{fig:g-cno} and \ref{fig:g-nneo}.
}
\label{fig:g-hehno}
\end{figure}

Figures \ref{fig:g-nneo} and \ref{fig:g-hehno} show the distributions of Ne/O vs. N/O and N/O vs. He/H, respectively.
Obviously, the distributions of values measured are much wider than those calculated. As the last paragraph discussed,
large observational errors and rough theoretical models can result in this mismatch.

\subsection{Isotopic ratios of CNO}
\begin{figure}
 \begin{center}
\includegraphics[totalheight=3.0in,width=2.2in,angle=-90]{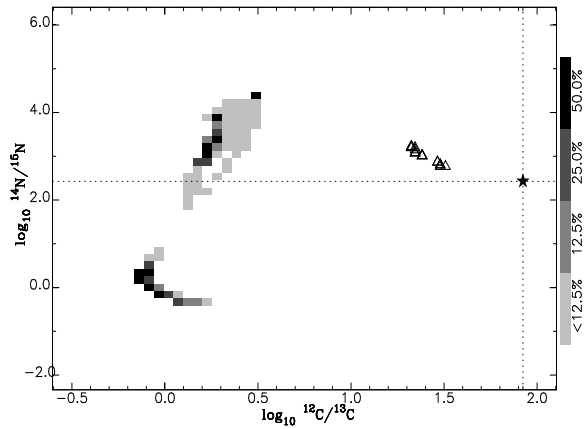}
 \end{center}
\caption{Gray-scale map of the isotopic ratios for nitrogen versus carbon for CO and ONeMg novae.
The top region represents novae with CO WD accretors
and the bottom region represents novae with ONeMg WD accretors.
The triangles represent the abundance values of AGB \citep {Halabi2015},
and the filled star is the abundance of the solar \citep {Anders1989}.
}
\label{fig:g-nc123}
\end{figure}

CNe are the major sources of the isotopes $^{13}$C, $^{15}$N and $^{17}$O \citep{Starrfield1997}.
Following \citet{Anders1993}, we also calculate the distributions of isotopic abundance ratios in novae.
Figure \ref{fig:g-nc123} displays the isotopic abundances ratios of \rm log $^{14}$N/$^{15}$N vs. \rm log $^{12}$C/$^{13}$C
in nova ejecta. The distributions cut into two regions.
The top region represents novae with CO WD accretors,
and the bottom region represents novae with ONeMg WD accretors, respectively.
Using a grid of evolutionary sequences of stars, \citet{Halabi2015} calculated the isotopic abundances ratios of \rm log $^{14}$N/$^{15}$N vs. \rm log $^{12}$C/$^{13}$C of some red giants, which are plotted by the triangles in Figure \ref{fig:g-nc123}.
It is evident that $^{12}$C/$^{13}$C in nova ejects is much lower than that in red giant. That is,
compared to red giants, novae can efficiently produce $^{13}$C. Similarly, novae with ONeMg WD can produce
more efficiently $^{14}$N.

\begin{figure}
 \begin{center}
\includegraphics[totalheight=3.0in,width=2.2in,angle=-90]{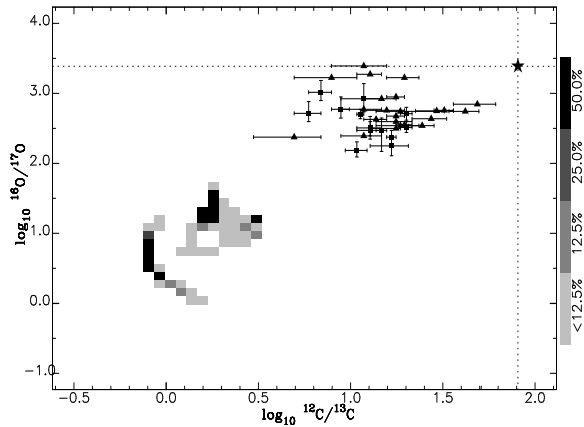}
 \end{center}
\caption{Gray-scale map of the isotopic ratios for oxygen versus carbon for CO and ONeMg novae.
The right above region represents novae with CO WD accretors
and the left bottom region represents novae with ONeMg WD accretors.
The filled squares and the filled triangles represent the observed values in AGB
in \citet {ElEid1994} and in \citet {Smith1990}, respectively.
We use the same notations as in Figure \ref{fig:g-nc123},
the filled star is the abundance of the solar \citep {Anders1989}.
}
\label{fig:g-oc167}
\end{figure}

\citet {ElEid1994} summarized the observed values of  isotopic abundance ratios of C, N and O
of 13 red giants. Using near-infrared and infrared spectra, \citet{Smith1990} estimated
the isotopic abundance ratios of C, N and O for red giants. These values are plotted by
the filled squares and the filled triangles in Figure \ref{fig:g-oc167}, respectively.
Obviously, compared to red giants, novae can efficiently produce $^{17}$O.

\subsection{The contribution of chemical abundances to the ISM}
As the last section discussed, novae can efficiently produce isotopic C, N and O.
We calculate annually contribution of isotopic C, N and O ,as well as the other isotopes, from novae to the ISM of the Galaxy, which
are listed in Table \ref{tab:phase}.
The first column gives the compositions in nova ejecta, column 2, column 3 and column 4 provide the annually average ejected masses of isotopes for CO novae, ONeMg novae and all of novae, respectively.
Hence, the average ejected mass of $^{13}$C  by novae is $3.2\times10^{-7} $ $ \rm M_\odot\ {\rm yr^{-1}}$,
the average yields of isotopes $^{15}$N and $^{17}$O are $1.1\times10^{-8} $
$ \rm M_\odot\ {\rm yr^{-1}}$ and $2.8\times10^{-7} $ $ \rm M_\odot\ {\rm yr^{-1}}$, respectively.
We can estimate that the contribution of $^{13}$C, $^{15}$N and $^{17}$O produced by novae is about 10$\%$, 5$\%$ and 20$\%$, respectively.
In short,
although the overall mass contribution to the ISM produced by novae is little,
novae are the main sources of the odd-numbered nuclei $^{13}$C,  $^{15}$N and $^{17}$O in the ISM.

\begin{table}
\footnotesize
\caption{
Compositions and masses of nova model ejecta.
}
\begin{center}
\begin{tabular}{lllllllllll} 
\hline\hline
Ejecta Composition \ \ &
\multicolumn{2}{c}{Ejecta Mass ($ \rm M_\odot\ {\rm yr^{-1}}$)} \\ \cline{2-4}
& CO Novae & ONeMg Novae  & All of Novae \\
\hline
       $^{1}$H \ \ &         $4.0\times10^{-5} $ \ \ &   $1.3\times10^{-7} $  \ \ &         $4.0\times10^{-5} $ \\
       $^{4}$He \ \ &        $2.0\times10^{-5} $ \ \ &   $8.6\times10^{-8} $  \ \ &        $2.0\times10^{-5} $\\
       $^{12}$C \ \ &        $5.8\times10^{-7} $ \ \ &   $8.6\times10^{-9} $  \ \ &        $5.9\times10^{-7} $\\
       $^{13}$C \ \ &        $3.1\times10^{-7} $ \ \ &   $1.1\times10^{-8} $   \ \ &        $3.2\times10^{-7} $\\
       $^{14}$N \ \ &        $7.3\times10^{-6} $ \ \ &   $1.1\times10^{-8} $   \ \ &        $7.3\times10^{-6} $  \\
       $^{15}$N \ \ &        $1.4\times10^{-9} $ \ \ &   $9.3\times10^{-9} $   \ \ &        $1.1\times10^{-8} $\\
       $^{16}$O \ \ &        $6.8\times10^{-6} $ \ \ &   $6.0\times10^{-8} $  \ \ &        $6.9\times10^{-6} $\\
       $^{17}$O \ \ &        $2.7\times10^{-7} $ \ \ &   $1.2\times10^{-8} $   \ \ &        $2.8\times10^{-7} $\\
       $^{20}$Ne \ \ &       $5.9\times10^{-8} $ \ \ &   $7.8\times10^{-8} $   \ \ &       $1.4\times10^{-7} $\\
       $^{22}$Ne \ \ &       $3.6\times10^{-7} $ \ \ &   $8.1\times10^{-10} $   \ \ &       $3.6\times10^{-7} $\\
\hline \hline
\end{tabular}
\end{center} \label{tab:phase}
\end{table}

\section{Conclusions}
The theoretical models of novae have been investigated by \citet{Yaron2005} and \citet{Jose1998}.
According to their models and using population synthesis technology,
we estimate that the occurrence rate of CNe in the Galaxy is about 54 $\rm yr^{-1}$, and the ejected mass by CNe is about
$2.7\times10^{-3}$ $ \rm M_\odot\ {\rm yr^{-1}}$.
The contributions of C, N and O elements to the ISM in the Galaxy are about $9.1\times10^{-7} $, $7.3\times10^{-6} $
and $7.1\times10^{-6} $ $ \rm M_\odot\ {\rm yr^{-1}}$, respectively. However, in the ejecta, the isotopic ratios of C, N and O
are higher about one order of magnitude than those in red giants.
Although the contribution of the ejecta produced by novae to the ISM is very little,
about 10$\%$, 5$\%$ and 20$\%$ of $^{13}$C, $^{15}$N and $^{17}$O
in the ISM of the Galaxy come from nova ejecta.
They are also required for understanding the CNO isotope evolution
and fairly essential and meaningful.
Unfortunately, our nova model is too simple to explain chemical abundances observed in CNe,
this means that there is still a long way to go in terms of understanding novae.


\section*{Acknowledgments}
This work was supported by
XinJiang Science Fund for Distinguished Young Scholars under Nos. 2014721015 and 2013721014, the National Natural Science Foundation
of China under Nos. 11473024, 11363005 and 11163005.

\bibliographystyle{apj}
\bibliography{lfeapj}


\label{lastpage}

\end{document}